\documentclass{article}

\PassOptionsToPackage{numbers, compress}{natbib}
 \usepackage[final]{nips_2017}
\usepackage[utf8]{inputenc} 
\usepackage[T1]{fontenc}    
\usepackage{hyperref}       
\usepackage{url}            
\usepackage{booktabs}       
 \usepackage{amsfonts}       
 \usepackage{nicefrac}       
 \usepackage{microtype}      
\usepackage{graphicx}
 \usepackage{color}
 \usepackage{float}
 \usepackage{subfigure}
 \usepackage{soul}
 \usepackage[table]{xcolor}
 \usepackage{amsmath}
\title{Time-Contrastive Learning Based DNN Bottleneck Features  for Text-Dependent Speaker Verification}

\author{ Achintya kr. Sarkar   \\
  Department of Electronic Systems\\
  Aalborg University, Denmark\\
  \texttt{akc@es.aau.dk} \\
  \And
  Zheng-Hua Tan\\
  Department of Electronic Systems\\
  Aalborg University, Denmark\\
  \texttt{zt@es.aau.dk} 
}

\begin{document}

\maketitle

\begin{abstract}
 In this paper, we present a time-contrastive learning (TCL) based bottleneck (BN) feature extraction method for speech signals with an  application to text-dependent (TD) speaker verification (SV). It is well-known that speech signals exhibit quasi-stationary behavior in and only in a short interval, and the TCL method aims to exploit this temporal structure. More specifically, it trains deep neural networks (DNNs) to discriminate temporal events obtained by uniformly segmenting speech signals, 
in contrast to existing DNN based BN feature extraction methods that train DNNs using labeled data to discriminate speakers or pass-phrases or phones or a combination of them. In the context of speaker verification, speech data of fixed pass-phrases are used for TCL-BN training, while the pass-phrases used for TCL-BN training are excluded from being used for SV, so that the learned features can be considered generic. The method is evaluated on the RedDots Challenge 2016 database. Experimental results show that TCL-BN is superior to the existing speaker and pass-phrase discriminant BN features and the Mel-frequency cepstral coefficient feature for text-dependent speaker verification.   
\end{abstract}

\section{Introduction}
\label{seg:introd}
Speaker verification (SV) aims to either accept or reject a person by his/her voice. It is broadly categorized into  text-dependent (TD)  and text-independent (TI) SV. In TD-SV, speakers are obliged to speak the same pass-phrases/sentences during both enrollment and test phases. In TI-SV, speakers are free to speak any pass-phrase/sentence during enrollment and test phases. As TD-SV maintains a matched phonetic content during both the training and test phrases, it outperforms TI-SV.

Due to the quasi-stationary behavior of speech within a short duration, short-time ceptral feature extraction is commonly deployed for speaker \cite{Kinnunen2010} and speech recognition \cite{Zheng-Hua2010}. Deep neural networks (DNNs) \cite{Hinton2012} have recently gained great attention from the  speech and speaker recognition community. 
In SV,  DNNs are generally  adopted either for  extracting the posteriori statistics \cite{McLaren2015,Kenny2014, Dahl2012} with respect to the pre-defined phonetic classes (called senones) for incorporating phonetic knowledge into i-vector extraction \cite{Deka_ieee2011} or for discriminative feature extraction, where a DNN is trained to discriminate speakers, pass-phrases, phonetic classes, phones or a combination of them. The  outputs of the DNN hidden layers are either directly used for  speaker characterization \emph{called d-vector} \cite{Variani2014} or projected onto a low dimensional space  \emph{called bottleneck (BN)} feature \cite{Fu2014,Yuan2015,Cong-Thanh2013,Ghalehjegh2015}. Many studies in literature have demonstrated \cite{Yuan2015,Cong-Thanh2013,Ghalehjegh2015} that BN feature based systems either perform  better than or provide complementary information to  conventional  short-time cepstral feature based speaker recognition.

A recent study \cite{Yuan2015} shows that the performance of TD-SV using BN features extracted  based on the discrimination of both speaker and pass-phrases is similar to that of features based on discrimination of either speakers or speaker+phone. Among them, augmentation of a cepstral feature with the speaker+pass-phrase discriminant BN feature yields the lowest error rates. However, all of these DNN BN feature extraction methods exploit label/supervision  information of data. The success of these systems highly  depends on the availability of well-labeled data. 

Inspired by the time contrastive learning (TCL) concept, a type of unsupervised feature learning used for classification of magnetoencephalography (MEG) data in \cite{Aapo2016}, we study the potential of TCL for speech signals. In  \cite{Aapo2016}, a TCL method is presented to classify few different states of brain that generally evolve over the time and can be measured by MEG signals. Speech and MEG signals, however, are quite different in nature, namely speech signals contain much richer information for which the tasks in hand often involve classification of much more classes.

In this paper, we explore the TCL concept for speech feature extraction. 
Two different strategies are considered. The first strategy randomly concatenates training utterances to form a single speech stream and then the stream is evenly partitioned into  segments, each of which is assumed to contain a single content belonging to a class. Let $N$ denotes the number of class in TCL. We take $N$ segments each time, and the data points within the $n^{th}$, $n\in\{1,2,...,N\}$, segment are assigned to class $n$. Then we take the next $N$ segments and assign data points the same way and so on. 
Afterwards, a DNN is trained to discriminate the data points. Finally, the output of a selected hidden layer is projected onto a low dimensional space to get  BN features. 
The second strategy is similar to the previous one with a key difference that \emph{each speech utterance} is uniformly divided into a number of segments (i.e. the number of classes in TCL) regardless of speakers and contents. 

The TCL-BN feature is experimentally compared with cepstral and existing BN features in TD-SV on the RedDots Challenge 2016 database consisting of short utterances. Experimental results show the superior performance of the TCL-BN feature. The Gaussian mixture model - universal background model (GMM-UBM) \cite{reynold00} technique is used  for SV since it is well-established that a GMM based classifier \cite{RSR2015,Delgado2016Asru} outperforms the i-vector \cite{Deka_ieee2011} method for SV with short utterances.  

\section{Cepstral and DNN bottleneck features}
{\bf Cepstral feature:} The Mel-frequency cepstral coefficients (MFCC) feature vectors (with RASTA \cite{Hermanksy94} filtering) are of $57$ dimensions consisting of static  $C_1$-$C_{19}$, $\Delta$ and $\Delta \Delta$ coefficients. The frame shift is $10$ ms and a $20$ ms Hamming window is used. An energy based voice activity detection is applied to remove the less energized frames. High energy frames are normalized to zero mean and unit variance at utterance level.\\
{\bf DNN bottleneck features:} Two DNN training approaches \cite{Yuan2015} are considered for BN feature extraction. In the first case, DNNs are trained to optimize a cross-entropy based objective function for discriminating speakers. 
In the second case, DNNs are trained to optimize two cross-entropy based objective functions simultaneously: one for discriminating speakers (\emph{spkr}) and the other for discriminating pass-phrases. There are two types of output nodes: one predicting speakers and the other predicting pass-phrases. Average of the two criteria is used as a final criterion in the DNN multi-task learning procedure \cite{CNTK}. The output of a DNN hidden layer at frame-level is then projected  onto a lower dimensional space \emph{called bottleneck features}, as illustrated in Fig.\ref{fig:sys1}.

\section{Time-contrastive learning concept and TCL based speech BN feature}
\begin{figure}[h]
  {\subfigure[\it]{ \hspace*{-0.0cm}\includegraphics[height=4.9cm,width=5.00cm]{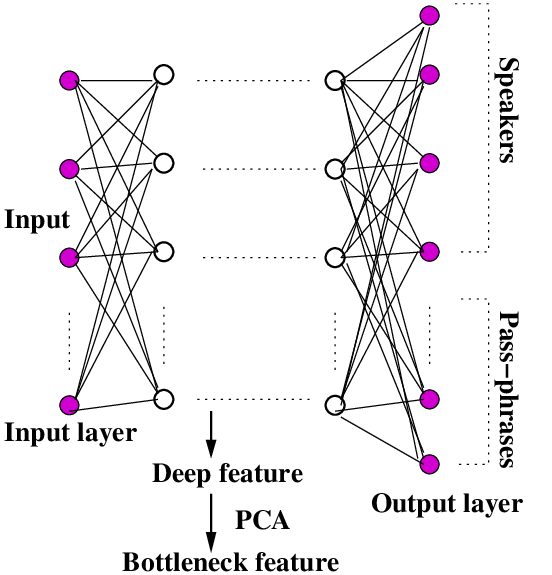}   \label{fig:sys1} } }
 \vline
  {\subfigure[\it]{\includegraphics[height=4.9cm,width=8.8cm]{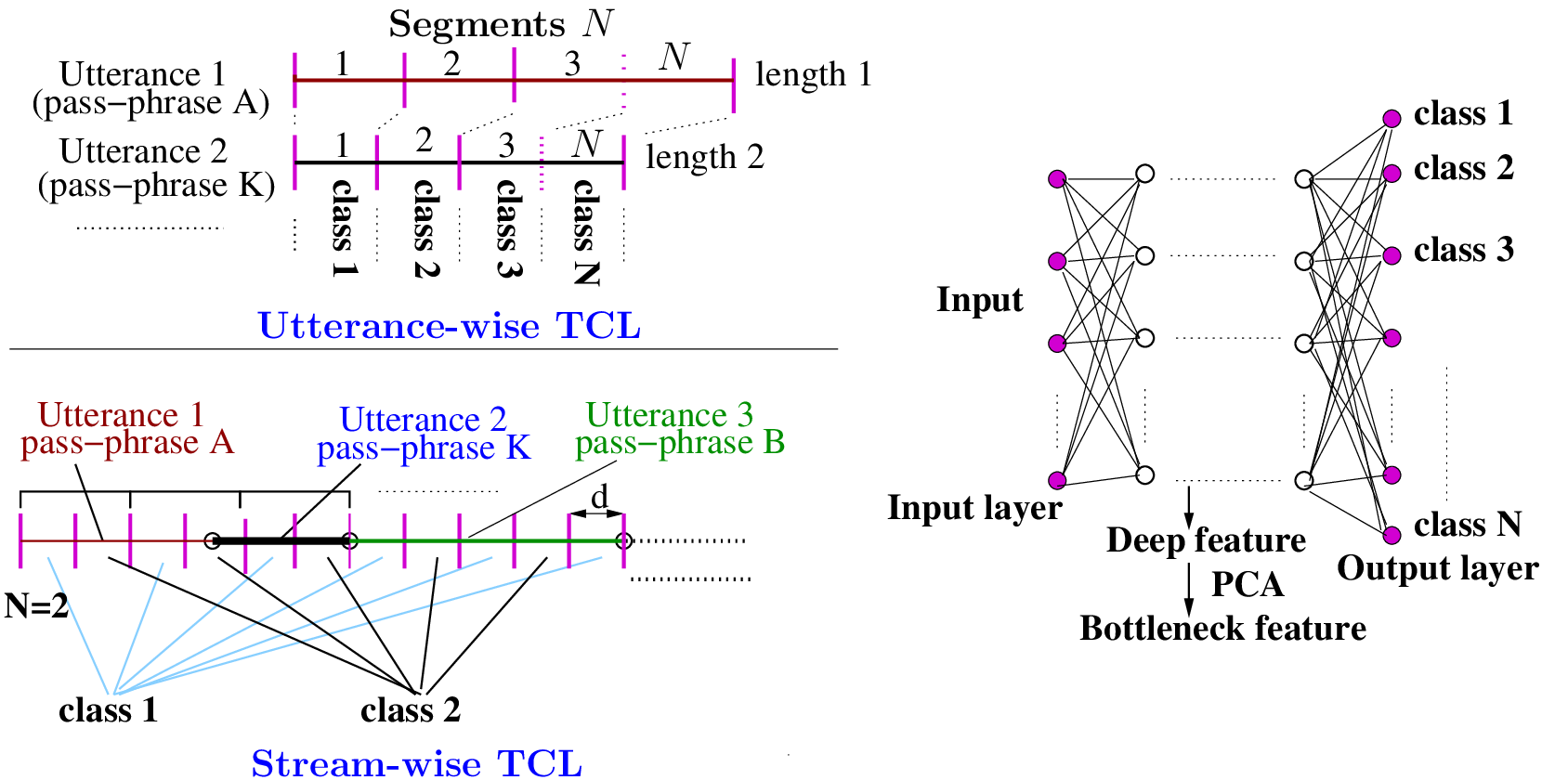}  \label{fig:sys2} }}
  \caption{\it Bottleneck feature extraction in (a) speakers + pass-phrases  (b) TCL}
  \vspace*{-0.3cm}
 \end{figure}

In the TCL  concept \cite{Aapo2016},  multivariate time series data $X$ are first divided into a number of uniform segments (say $N$), and then \emph{data-points} within a particular segment are assigned to one class label,\\
\begin{equation}
\underbrace{(x_1, ..., x_M)}_\text{class $1$}, \ldots, \underbrace{(x_{iM+1}, ..., x_{iM+M})}_\text{$\ldots$}, \ldots, \underbrace{(x_{(N-1)M+1}, ..., x_{NM})}_\text{class $N$}
\end{equation}
where $i$ and $M$ indicate the segment index and the number of data points within a segment, respectively. 
Finally, a DNN is trained to classify the data points. In \cite{Aapo2016}, the output of the last hidden layer is  used as a feature to classify different states of brain  using  magnetoencephalography (MEG) data. We adopt this concept and devise two strategies to extract speech features as follows. \\
{\bf Stream-wise TCL (sTCL):} Speech utterances of training data are randomly concatenated to form a single speech stream which is then partitioned into segments of $d=6$ frames each in order to capture short-time speech events. For a $N$ number of classes in TCL,  $N$ segments are taken at a time, and the data points within the $n^{th}$, $n\in\{1,2,...,N\}$, segment are assigned to class $n$. Then, the next consecutive $N$ segments are taken and assign data points the same way. The process continues until we finish assigning all data points in the stream. Afterwards, a DNN is trained to discriminate the data points assigned to $N$ different classes. Finally, the output of one of the DNN hidden layers is projected onto a low dimensional space to get  BN features for TD speaker verification.\\
{\bf Utterance-wise TCL (uTCL):} 
This method considers \emph{each training utterance separately} and then uniformly divides it into  $N$ segments. Class assignment for data points and DNN training are done the same way as in the previous method. The motivation here is to segment  each utterance the same way, namely the first segment is the beginning of an utterance and the last the end of it. This consistency in segmentation is expected to be beneficial in particular when there are utterances of the same textual content, e.g. training data of fixed pass-phrases. As there are textually-repeating utterances, this can be regarded as weak supervision. Note that the pass-phrases appeared in the training data are excluded from evaluation, so the learned feature is not phrase-specific. 
The TCL-BN feature extraction methods are illustrated in Fig.\ref{fig:sys2}.

\section{Experiments}
For both existing and TCL based BN feature extraction, DNNs are trained  using the RSR2015 \cite{RSR2015} database. Data used are  $\approx 72764$ utterances recorded on $9$ sessions for $27$ pass-phrases from $300$ non-target speakers ($157$ male, $143$ female).  All DNNs are $7$ layer feed-forward networks and are trained using the same learning rate and the same number of epochs. Each hidden layer consists of $1024$ sigmoid units. 
The input layer is of $627$ dimensions, based on $57$ dimensional MFCC features with a context window of $11$ frames (i.e. $5$ frames left, current frame, $5$ frames right). In existing BN, the speaker-discriminant DNN consists of an output layer equal to the number of speakers, i.e. $300$ nodes, while the speaker+pass-phrase discriminant DNN consists of $327$ output nodes ($300$ speakers and $27$ pass-phrases).
To obtain the final BN feature, the output from a chosen hidden layer, a $1024$ dimensional deep feature, is projected onto a $57$ dimensional space to align with the dimension of MFCC feature for a fair comparison. Deep features are normalized to zero mean and unit variance at utterance level before using principle component analysis (PCA) for dimension reduction.

Experiments for text-dependent speaker verification are conducted on a different database, which is Part I (male speakers) of the RedDots database \cite{RedDots} as per protocol. 
Three types of non-target trials are available for  evaluation. For details about the  database and number of trials refer to \cite{RedDots}.    

Gender-independent GMM-UBM ($512$ mixtures, having diagonal covariance matrix) is trained using non-target speakers ($438$ male, $192$ female) data ($6300$ utterances) from the TIMIT database \cite{Timit}. The GMM-UBM training data are reused for the PCA.
Speaker models are derived from the GMM-UBM (updating mean vector of mixtures) with maximum a posteriori (MAP) adaptation using their respective training data.
In the test phase, a test utterance $Y=\{\textbf{y}_1,\textbf{y}_2, \ldots, \textbf{y}_T\}$ is scored against the target specific model (obtained in training) $\lambda_r$ and GMM-UBM $\lambda_{ubm}$. 
Finally, a log likelihood ratio (LLR)  value is calculated using the scores between the two models $LLR(Y) = \frac{1}{T} \sum_{t=1}^{T}\{\log \;p(\textbf{y}_t|\lambda_r) - \log\; p(\textbf{y}_t|\lambda_{ubm})\}$, $\textbf{y}_t$ denotes the $t^{th}$ frame/feature vector.
Three iterations (with value of relevance factor $10.0$) are used in MAP. 
System performance is evaluated in terms of equal error rate (EER) and minimum detection cost function (minDCF) \cite{DET97}.

\section{Results and discussion}
Table \ref{table:table0} shows the effect of number of classes ($N$) in TCL-BNs (for the $L2$ layer) on TD-SV. It is observed  that sTCL and uTCL give lower TD-SV error rates when $N=15$ and $N=10$, respectively, and hence those values are experimentally chosen to represent their optimal performance in this setting. In this work, TCL-BN features are extracted from the $L2$ output layer of DNN, which gives slightly better performance than from $L4$.
The uTCL approach outperforms sTCL, which could be explained by the factor that uTCL takes advantage of consistently segmenting utterances of the same textual content in contrast to sTCL that concatenates all utterances together before segmentation.
Table \ref{table:table1} compares the performance of TD-SV on different feature sets (BN feature for the best DNN layer output in the respective systems: L4 in existing and L2 for TCL) for different non-targets. 

\begin{table}[h]
\caption{\it Effect of number of classes ($N$) in TCL-BNs (L2 layer output of DNN) on TD-SV. The average EER is calculated over the system performance for different types of non-target (see Table\ref{table:table1})}
\vspace*{-0.1cm}
\begin{center}
\begin{tabular}{|lllllll|}\cline{1-7}
Feature set      &\multicolumn{6}{c|}{Number of classes ($N$) in TCL [\% average EER]}\\
                 && &5    &  10  & 15   & 20 \\ \hline
sTCL-BN          && &2.91  &2.88  &{\bf 2.83}  &2.86\\
uTCL-BN          && &1.98  &{\bf 1.89}  & 2.96 &2.95 \\ \hline             
\end{tabular}
\end{center}
\label{table:table0}
\end{table}
\begin{table}[h]
\caption{\it Comparing performance of TD-SV for different features on RedDots m\_part\_01.}
\begin{center}
\begin{tabular}{|llll|l|}\cline{1-5}
  Feature set     &\multicolumn{3}{c}{Non-target type [\%EER/(minDCF$\times$ 100)]} & Avg. (EER\\
            & Target-wrong      & Impostor-correct & Impostor-wrong            & /minDCF) \\ \hline 
  MFCC              & 5.12/2.176      & 3.33/1.401        & 1.14/0.474 & 3.19/1.350 \\ 
 BN-spkr            & 4.59/1.654      & 3.05/1.355        &1.11/0.380  & 2.91/1.130\\ 
 BN-spkr+phrases        & 4.53/1.644      &3.07/{\bf 1.348}         &1.17/0.385  & 2.92/1.125 \\ 
 sTCL-BN  [N=15]           &4.33/1.662 & {\bf 3.02}/1.384  & 1.14/0.391 & 2.83/1.145  \\
 uTCL-BN  [N=10]    &{\bf 1.88/0.654}       &3.14/1.444         &{\bf 0.64/0.195}  &{\bf 1.89/0.764}  \\ \hline             
\end{tabular}
\end{center}
\label{table:table1}
\end{table}
Table \ref{table:table1} shows TD-SV results for different features. Similarly to \cite{Yuan2015}, existing BN features (for the $L4$ layer) show better TD-SV performance than the cepstral feature, and the two existing BN features show performance close to each other. TCL-BN shows  lower average error rates for TD-SV compared with the cepstral and  existing  BN features. Specially, uTCL-BN shows significant reduction of error rates for the target- and impostors-wrong cases, which could be due to its ability to capture phonetic discriminate information. Overall, the TCL based features are effective. Moreover, it is observed in \cite{Yuan2015}  that the TD-SV performance of BN features extracted from DNNs trained to discriminate speaker and triphone state labels (from speech recognition with a supervised mode) is very similar to those of \emph{BN-spkr} and \emph{BN-spkr+phrases}. However, speech recognition systems need annotation/labeled data for training. On the other hand, the TCL method does not use any  speaker/pass-phrase/phonetic specific label  information. 
\section{Conclusion}
In this paper, we explored the time-contrastive learning (TCL) concept for training DNN based BN features for text-dependent speaker verification (TD-SV), in which DNNs are trained to discriminate the temporal events across a speech signal. This is realized by uniformly segmenting the speech signal into a number of segments and assigning the same label to all speech frames in one segment but different labels to different segments. DNNs are then trained to discriminate data across the different time segments without any speaker or phonetic label in contrast to the existing DNN BN feature extraction approaches. Experimental results confirmed the effectiveness of the proposed methods.  
\subsubsection*{Acknowledgement}
This work is supported by the iSocioBot project, funded by the Danish Council for Independent
 Research - Technology and Production Sciences (\#1335-00162).
\bibliographystyle{IEEEbib}
\bibliography{strings,References}

\end{document}